\documentclass{aastex6}

\begin{document}


\title{A new look at the integrated radio/microwave continuum spectrum of Galactic supernova remnant IC 443}


\author{D. Oni{\' c}\altaffilmark{1}, D. Uro{\v s}evi{\' c}\altaffilmark{1,2}, and D. Leahy\altaffilmark{3}}



\altaffiltext{1}{Department of Astronomy, Faculty of Mathematics, University of Belgrade, Serbia; donic@matf.bg.ac.rs}
\altaffiltext{2}{Isaac Newton Institute of Chile, Yugoslavia Branch}
\altaffiltext{3}{Department of Physics and Astronomy, The University of Calgary, Canada}

\begin{abstract}
Recent observations of the microwave sky, by the space telescopes such as \textit{WMAP} and \textit{Planck}, have opened a new window into the analysis 
of continuum emission from supernova remnants (SNRs). In this paper, different emission models that can explain the characteristic shape of presently 
known integrated radio/microwave continuum spectrum of the Galactic SNR IC 443 are tested and discussed. In particular, the possibility that the 
slight bump in the integrated continuum of this remnant around 20 -- 70~GHz is genuine and that can be explained by the contribution of additional emission 
mechanism such as of spinning dust is emphasized. We find that adding a spinning dust component to the emission model improves the fit of the 
integrated spectrum of this SNR while, at the same time preserves the physically probable parameter values. Finally, models that include 
the high-frequency synchrotron bending of the IC 443 radio to microwave continuum are favored. 
\end{abstract}




\section{Introduction} \label{sec:intro}

The radio-continuum spectra of SNRs are generally shaped by the (non-thermal) synchrotron radiation. In an external magnetic field 
$B~\mathrm{[\mu G]}$, an electron of energy $E~\mathrm{[GeV]}$ radiates its peak power at a frequency $\nu~\mathrm{[GHz]}$, 
following the particular relation: $E=14.7\sqrt{\nu/B}$ (Reynolds 2008). That leads to the conclusion that for the standard value 
of the mean Galactic magnetic field (order of $\mathrm{\mu G}$), GeV electrons are responsible for the observed synchrotron emission 
in GHz range. The most probable and efficient mechanism for a production of high energy particle ensemble in SNRs is the diffusive 
shock acceleration (DSA; Bell 1978; Blandford \& Ostriker 1978). It produces the non-thermal ensemble which in the simplest, 
test-particle case has a power-law energy distribution (see Uro\v sevi\'c 2014, and references therein for a review). The analysis 
of the integrated radio to microwave continuum of SNRs (at broad range of frequencies) is important as possible deviations from the 
theoretical predictions can give us the new insights into physics behind the observed radiation.

Generally, verification of several theoretical models (e.g., non-linear particle acceleration in young SNRs, significant 
thermal bremsstrahlung emission from the SNRs expanding in the dense environment, models of dust emission linked to the SNRs, 
etc) rely, particularly on a good knowledge of the high-frequency part of the radio as well as the microwave continuum 
of SNRs (Reynolds \& Ellison 1992; Scaife et~al.\@ 2007; Oni\'c et~al.\@ 2012; Oni\' c \& Uro{\v s}evi{\' c} 2015; Oni\' c 2015; 
G\'{e}nova-Santos et~al.\@ 2016). The ground based radio-observations of SNRs at frequencies higher than around 10 GHz generally 
suffer from the transparency issues due to the existence of Earth's atmosphere. In that sense, the high altitude and/or space 
observatories are needed for the analysis of high-frequency radio to microwave spectral range.

Recently, the observations from the microwave survey of Galactic SNRs made by {\it Planck}\footnote{{\it Planck} 
(http://www.esa.int/Planck) is a project of the European Space Agency (ESA) with instruments provided by two scientific 
consortia funded by ESA member states, with contributions from NASA (USA) and telescope reflectors provided by a 
collaboration between ESA and a scientific consortium led and funded by Denmark.} became available 
(Planck Collaboration Int.\@ XXXI 2016). {\it Planck} has taken the all-sky observations in nine frequency bands between 30 -- 857~GHz. 
The Low Frequency Instrument on board {\it Planck} covered the 30, 44 and 70~GHz bands with angular 
resolutions of around $33\arcmin$, $24\arcmin$ and $14\arcmin$, respectively (Mandolesi et~al.\@ 2010; Planck Collaboration VI 2016). 
The High Frequency Instrument, however, covered the 100, 143, 217, 353, 545 and 857~GHz bands with an angular resolution ranging from 
around $9\farcm{}9$ to $4\farcm{}4$ (Planck HFI Core Team 2011; Planck Collaboration VIII 2016). For a comparison, the angular resolution of 
\textit{Wilkinson Microwave Anisotropy Probe} (\textit{WMAP}) at 23, 33, 41, 61 and 94~GHz is around 
$53\arcmin$, $40\arcmin$, $31\arcmin$, $21\arcmin$ and $13\arcmin$, respectively (Bennett et~al.\@ 2003). That makes 
\textit{WMAP} unsuitable for the analysis of most Galactic SNRs (see Green 2014). Only a few Galactic SNRs, with large enough 
angular dimensions were detected with \textit{WMAP} (Cas A, Puppis A, HB 21, W44 region; Weiland et~al.\@ 2011; Hewitt et~al.\@ 2012; 
Pivato et~al.\@ 2013; Irfan et~al.\@ 2015; G\'{e}nova-Santos et~al.\@ 2016).

In this paper, the integrated continuum radio to infrared spectrum of SNR IC 443 is analyzed testing the different emission 
models that can be responsible for its particular shape. Section 2 is dedicated to the main characteristics of the remnant 
while Section 3 focuses on the properties of its radio/microwave continuum. Section 4 discusses different theoretical emission 
models that are appropriate in the analysis of this SNR. In particular, the hypothesis that the bump in the integrated continuum of 
the SNR IC 443 around 20 -- 70~GHz is genuine and that can be explained by the contribution of additional emission such as of 
spinning dust is analyzed in Section 5 and further discussed in Section 6. The final section summarizes the main results of the 
analysis.

\section{SNR IC 443}
\label{sec:SNR}

The SNR IC 443 (G189.1+3.0, 3C 157), also known as the Jellyfish Nebula in visible sky, has a rather large angular size 
among the majority of Galactic SNRs detected in radio domain ($45\arcmin$; Green 2014). In addition to its apparent 
dimensions, the rough proximity to the Galactic anti-center location, leaves this remnant relatively well isolated from 
the confusing effects normally complicating studies of the inner Galactic SNRs (Castelletti et~al.\@ 2011; 
Ohnishi et~al.\@ 2014). A distance of 1.5 kpc (corresponding the diameter of 20 pc) is adopted in most of the papers related 
to IC 443 although it is still debated (Fesen 1984; Welsh \& Sallmen 2003). In fact, this SNR is also probably in a physical 
connection with the \mbox{H\,{\sc ii}} region SH 2-249 located north of IC 443 which leads to a bit longer distance of 
1.5 -- 2~kpc (Reich et~al.\@ 2003; Gao et~al.\@ 2011). Roughly estimated age of this SNR is of 3000 to 30,000~yr 
(Petre et~al.\@ 1988; Olbert et~al.\@ 2001; Leahy 2004) or more precise, 20,000~yr as suggested in Lee et~al.\@ (2008).

This SNR has been extensively observed and analyzed throughout the whole electromagnetic spectrum from radio-frequencies to $\gamma$-rays. 
IC 443 shows a limb-brightened morphology at optical, infrared, and radio-wavelengths, while it manifests a centrally peaked 
morphology of the thermal bremsstrahlung origin in the X-rays (Kokusho et~al.\@ 2013a, and references therein). As a consequence, 
it has been classified as the mixed-morphology or thermal-composite SNR (Rho \& Petre 1998). Several robust evidences for the 
presence of recombining plasma are based on the X-ray emission analysis of this remnant (Kawasaki et~al.\@ 2002; Yamaguchi et~al.\@ 2009; 
Ohnishi et~al.\@ 2014).

Based on the radio and X-ray morphology and spectral analysis, as well as the radio polarization properties, Olbert et~al.\@ (2001) found 
that the non-thermal, X-ray source, located in the southern portion of IC 443, is a synchrotron nebula powered by a central 
compact point source (with soft X-ray thermal spectrum which is consistent with the emission from the surface of a neutron 
star; Bocchino \& Bykov 2001). They also suggested that it is physically associated with IC 443. The existence of a pulsar wind nebula 
(Swartz et~al.\@ 2015) and a metal-rich X-ray plasma (Troja et~al.\@ 2008) would indicate that IC 443 is of a core-collapse origin 
from a massive progenitor star (Kokusho et~al.\@ 2013a), probably a $15-19\ \mathcal{M}_{\odot}$ B0 star (Su et~al.\@ 2014). 
Still, there remains some doubt as to the physical association of the pulsar wind nebula and the SNR IC 443 (Leahy 2004; 
Gaensler et~al.\@ 2006). So far, no pulsations were detected from the neutron star.

SNR IC 443 is evolving in a rich and complex interstellar region. It interacts with both low and high-density material, 
which strongly affects the evolution of SNR. It is one of the prototypical cases of SNRs impacting dense interstellar molecular gas. 
In fact, it is associated with a dense giant molecular cloud (MC) near the Gem OB1 association (Cornett, Chin \& Knapp 1977; Humphreys 1978; 
Heiles 1984). The SNR/MC interaction in IC 443 is very well studied (Hoffman et~al.\@ 2003; Shinn et~al.\@ 2011; 
Su et~al.\@ 2014; Kilpatrick, Bieging \& Rieke 2016, and references therein). Actually, the physical conditions toward this 
SNR are very well suited for all known shock interaction tracers that are traditionally used to identify SNR/MC interactions (broadened 
CO emission, OH 1720 MHz maser, HCO$^{+}$, HCN, $\mathrm{H_{2}}$ emission, etc). Furthermore, detection of the characteristic 
pion-decay feature in $\gamma$-ray spectrum provides a direct evidence that cosmic-ray protons (which penetrate into high 
density MCs) are indeed accelerated in SNRs (Abdo et~al.\@ 2010; Tavani et~al.\@ 2010; Ackermann et~al.\@ 2013; 
Tang \& Chevalier 2014; Tang \& Chevalier 2015).

Oliva et~al.\@ (1999) found that IC 443 is characterized by prominent line -- emitting filaments and relatively strong 
\textit{IRAS} (\textit{Infrared Astronomical Satellite}) 12 and 25 $\mathrm{\mu{m}}$ emission with most of the flux accounted 
for by ionized line emission (mainly of [\mbox{Ne\,{\sc ii}}] and [\mbox{Fe\,{\sc ii}}]). They also noted possible contribution 
of $\mathrm{H_{2}}$ lines to the \textit{IRAS} 12 $\mathrm{\mu{m}}$ flux from the southern rim of IC 443. In other words, 
Oliva et~al.\@ (1999) suggested that lines account for most of the \textit{IRAS} 12 and 25 $\mathrm{\mu{m}}$ emission from the 
line -- emitting filaments of IC 443. That is in contrast with the common interpretation involving thermal radiation from very 
small grains stochastically heated by collisions with the hot plasma behind the shock front. In addition, Kokusho et~al.\@ (2013a) 
have found that the [\mbox{Fe\,{\sc ii}}] line emission is enhanced relative to the thermal emission from the warm dust in the central 
region of IC 443. They concluded that ionized Fe, in that region, is probably mostly of the interstellar, rather than of ejecta origin. 
Kokusho et~al.\@ (2013b) emphasized the possibility that the most of Fe atoms are in fact contained in the deepest cores of dust grains, 
or that there is a population of Fe-rich dust which is relatively tough against sputtering.

It is worth mentioning that Hezareh et~al.\@ (2013) reported detection of non-Zeeman circular polarization and linear 
polarization levels of up to 1\% in the CO rotational spectral line emission in a shocked molecular clump around the SNR IC 443. 
They concluded that the non-Zeeman CO circular polarization is most probably due to a linear-to-circular polarization conversion, 
consistent with a physical model based on anisotropic resonant scattering of Houde et~al.\@ (2013). In fact, it is proposed that 
background, linearly polarized CO emission interacts with similar foreground molecules aligned with the ambient magnetic 
field and scatters at a transition frequency. Actually, the difference in phase shift between the orthogonally polarized components 
of this scattered emission can cause a transformation of linear to circular polarization. Furthermore, Koo et~al.\@ (2010) 
measured the Zeeman splitting of the \mbox{H\,{\sc i}} 21 cm emission line from shocked atomic gas in IC 443 and derived 
an upper limit of $B_{\parallel}=100-150\ \mathrm{\mu G}$ on the strength of the line-of-sight magnetic field component. They proposed that 
either the magnetic field is roughly random within the telescope beam due to inhomogeneities in preshock gas and/or various 
hydrodynamic instabilities, or alternatively, the preshock density may be low, much lower than that mean density of molecular 
clouds. Koo et~al.\@ (2010) emphasized that the latter is possible if the molecular cloud that the SNR is interacting with is clumpy 
and the high-velocity \mbox{H\,{\sc i}} emission is from shocked diffuse inter-clump medium. Finally, as the radio spectral index of this remnant 
(see Eq.\@ 1 and Section 3) is less than the characteristic value of 0.5, we can not use the standard method, i.e., the equipartition 
calculation (see Arbutina et~al.\@ 2012; 2013 for more details) to estimate the magnetic field strength.

\section{The radio/microwave continuum of SNR IC 443}
\label{sec:rmcon}

Two main sub-shells (shells A and B), with markedly different radial intensity distributions, make up the majority 
of the SNR IC 443 in radio continuum emission (Braun \& Strom 1986; Reich et~al.\@ 2003; Leahy 2004). They appear to be connected, 
roughly spherical, half-shells of radio synchrotron emission, which are centered at different spatial positions. Together, they 
define the usually assumed boundaries of the SNR. Shell A appears brighter, limb-brightened, and is coincident with the 
molecular shock tracers along its southern boundary and across the center of the SNR as a whole (Kilpatrick et~al.\@ 2016). 
Shell B is, however, dimmer and has predominantly uniform surface brightness with some enhancement which position coincides with 
the optical filaments (Lee et~al.\@ 2008). The radius of shell A is around $19\arcmin$ or 8.3~pc and of shell B is around $29\arcmin$ 
or 12.7~pc, for the assumed SNR distance of 1.5~kpc (Leahy 2004). The nature of third, larger, incomplete and faint shell 
extending beyond the northeast periphery of the remnant is still not clear enough although. Still, it was proposed to be a different SNR (G189.6+3.3), 
that overlaps with IC 443 (Asaoka \& Aschenbach 1994; Leahy 2004; Castelletti et~al.\@ 2011).

The radio morphology of IC 443 described above is consistent with the scenario whereby the western part of the remnant 
has actually broken out into a rarefied medium (Lee et~al.\@ 2008). Actually, it is proposed that the SNR has probably 
been formed inside a dense medium (possibly evolved inside the preexisting wind-blown bubble) and then broke out to the 
adjacent rarefied medium. In addition, Su et~al.\@ (2014) reported infrared detection of sixty-two young stellar object 
(YSO) candidates (disk-bearing young stars: 24 of Class I and 38 of Class II), mainly concentrated along the boundary 
of the remnant's bright radio shell (but absent in the southwestern breakout portion of the SNR), and suggested that 
they are likely to be triggered by the stellar wind from the massive progenitor of SNR IC 443.

The integrated radio-continuum spectrum of SNRs (integrated flux density vs.\@ frequency) is generally represented by a simple power-law 
that arises from the synchrotron emission of electrons accelerated by a DSA mechanism: \begin{equation}
S_{\nu}^{\mathrm{syn}}\propto\nu^{-\alpha},                     
\end{equation}
where $\alpha$ is the radio spectral index.

The intensity of radio synchrotron emission depends mainly on the energy of non-thermal electrons and the magnetic field strength. 
Generally, for the radiative shocks (usually present in mixed-morphology SNRs), the compression factors are large, giving rise 
to the strongly compressed magnetic fields, and increased cosmic-ray electron densities. For steady radiative shocks propagating 
through a uniform medium of the certain number density of hydrogen nuclei of preshock gas $n_{0}\ [10\ \mathrm{cm^{-3}}]$, the 
strength of the magnetic field $B_{\mathrm{max}}\ [\mu\mathrm{G}]$ normal to the shock velocity $v_{\mathrm{s}}\ [100\ \mathrm{km\ s^{-1}}]$ 
reaches to $B_{\mathrm{max}}\approx240\ v_{\mathrm{s}}\ \sqrt{n_{0}}$ (Chevalier 1974). As a result, the radio synchrotron emission from 
dense SNR shells is enhanced (van der Laan mechanism; van der Laan 1962). In fact, the emitting electrons may be either those accelerated 
by the SNR, or are simply ambient relativistic electrons swept up by the radiative shocks (Vink 2012). Of course, the radio emission may 
be additionally enhanced by the presence of secondary electrons/positrons, i.e., the electrons/positrons products left over from the decay 
of charged pions, created due to cosmic-ray nuclei colliding with the background plasma (Uchiyama et~al.\@ 2010). Finally, the models that 
assume the energy spectrum of the non-thermal electrons that is shaped by the joint action of first- and second-order Fermi acceleration 
(Ostrowski 1999) in a turbulent plasma with substantial Coulomb losses were also proposed and tested for the case of SNR IC 443 
(see Bykov et~al.\@ 2000 for more details).

This remnant also exhibits a turnover in its integrated radio-continuum spectrum at the lowest frequencies (below around 30 MHz). 
Contrary to the expectation that such absorption arises from unrelated low-density \mbox{H\,{\sc ii}} regions (or their 
envelopes) along the line of sight, Castelletti et~al.\@ (2011) proposed that in this case the absorbing medium is directly 
linked to the SNR itself. The evidence for a similar situation has also been observed in the case of 3C 391 (Brogan et~al.\@ 2005). 
Castelletti et~al.\@ (2011) reported the excellent correspondence between the observed eastern radio flattest spectrum region 
(spectral index actually vary across the SNR) and near infrared ionic lines which strongly suggests that the passage of a fast, 
dissociating J-type shock across the interacting molecular cloud actually dissociated the molecules and ionized the gas. 
They therefore concluded that such a collisional ionization is responsible for the thermal absorbing electrons that produce 
the peculiar very flat spectrum areas observed all along the eastern border of IC 443. This is in agreement with the model 
proposed by Rho et~al.\@ (2001) in which the infrared emission from the ionized species in the east bright radio limb of 
IC 443 comes from shattered dust produced by a fast dissociating J-type shock. Assuming an electron temperature in a range 
between 8000 -- 12,000~K (consistent with infrared analysis), Castelletti et~al.\@ (2011) estimated emission measure ($EM$) 
of $(2.8-5.0)\times10^{3}\ \mathrm{cm^{-6}\ pc}$ for the region of the strongest thermal absorption (eastern rim).

The flux density of, already mentioned, pulsar wind nebula in IC 443 at 1.42~GHz is of $0.20\pm0.04\ \mathrm{Jy}$ while the 
integrated flux density of the SNR is around 140~Jy at the same continuum frequency (Castelletti et~al.\@ 2011).

Finally, Planck Collaboration Int.\@ XXXI (2016) reported detection of the SNR IC 443 at all nine {\it Planck} frequencies and 
suggested that the spectral energy distribution (SED) across radio and microwave frequencies can be reasonably approximated by 
a combination of synchrotron and dust emission. They noted that the synchrotron emission roughly follows a power-law with 
radio spectral index of 0.36 from radio frequencies above 30~MHz up to 40~GHz, after which the spectral index steepens to 1.56. 
It is proposed that the decrease in flux density could be due to a break in the synchrotron power-law from the injection 
mechanism of the energetic particles, or due to cooling losses by the energetic particles. The higher-frequency emission was 
found to be, most likely due to dust grains that survive the shock. No indications of significant thermal bremsstrahlung emission 
from the SNR have been found by Planck Collaboration Int.\@ XXXI (2016) in contrast to the results of Oni\'c et~al.\@ (2012).

However, there is a slight bump seen in the, presently known, radio to microwave integrated continuum spectrum of this SNR 
around 20 -- 70~GHz (see Figs.\@ 2 and 3). That can be an indication of the significant presence of some other emission 
mechanism (such as of spinning dust emission) which is discussed bellow.

\section{The emission models}
\label{sec:models}

To account for the low-frequency spectral turnover, due to the thermal (free-free) absorption, and possible thermal bremsstrahlung 
emission, one can, generally, assume several models. On such model assumes that thermal absorption and emission actually originate 
from the same volume of space, just ahead of the region that emits synchrotron radiation. If the frequencies are given in GHz, 
we can write: \begin{equation} S_{\nu}^{\mathrm{M1}}=S_{\nu}^{\mathrm{syn}}\exp\left[{-\tau_{\nu}}\right]+S^{\mathrm{ff}}\nu^{2}\left(1-\exp\left[{-\tau_{\nu}}\right]\right), 
\end{equation} where $\tau_{\nu}\propto\nu^{-2.1}$ is an optical depth, $S_{\nu}^{\mathrm{syn}}$ is given by Eq.\@ 1 and 
$S^{\mathrm{ff}}$ corresponds to the thermal (free-free) flux density at 1~GHz. On the other hand, if we assume that both, 
thermal absorption and emission, as well as the synchrotron radiation, are coming from the same region, we have: \begin{equation} 
S_{\nu}^{\mathrm{M2}}=\left(S_{\nu}^{\mathrm{syn}}/\tau_{\nu}+S^{\mathrm{ff}}\nu^{2}\right)\left(1-\exp\left[{-\tau_{\nu}}\right]\right). 
\end{equation} In order to estimate the initial value of the parameter $S^{\mathrm{ff}}$ for the least-square fits (see Section 5), in the case of SNR IC 443, 
one can use the electron temperatures derived in Castelletti et~al.\@ (2011) and the value of $10^{-4}\ \mathrm{sr}$, for the source 
solid angle $\Omega_{\mathrm S}$ (Green 2014), as $S^{\mathrm{ff}}=2k_{\mathrm{b}}T_{\mathrm{e}}\Omega_{\mathrm{S}}/c^{2}$, where $k_{\mathrm{b}}$ 
and $c$ are Boltzmann constant and speed of light, respectively.

Surely, one should bear in mind that SNRs are in essence 3D structures which is the main drawback of these models. Also, the possibility of 
synchrotron self-absorption is not discussed here. It is generally very hard to discriminate between thermal absorption effects (in an optically 
thick region, the spectral index is $\alpha=2$) and synchrotron self-absorption ($\alpha=2.5$). However, Castelletti et~al.\@ (2011) showed 
that the main effect responsible for the low-frequency spectral turnover, in the case of SNR IC 443, comes from the thermal absorption that is 
associated with this SNR.

Thermal dust emission, which dominates continuum spectrum of IC 443 above 140~GHz, is usually very well described by the 
simple model represented by a modified black-body relation: \begin{equation} 
S_{\nu}^{\mathrm{Td}}\propto\nu^{\beta_{\rm d}}B_{\nu}(T_{\rm d}),\end{equation} where $B_{\nu}(T_{\rm d})$ is standard Planck, black-body, function 
for dust at temperature $T_{\rm d}$ and $\beta_{\rm d}$ is the (dust) emissivity index, usually between values of 0 -- 2 
(Blain et~al.\@ 2002; Planck Collaboration XI 2014). Planck Collaboration Int.\@ XXXI (2016) used {\it Planck's} data and fitted 
thermal dust emission with a one temperature model. They estimated a dust temperature of 16~K with emissivity index of 1.5, but as they 
noted, the precise values are not unique and require combination with infrared data and multiple temperature components given the complex 
mixture of dust in molecular, atomic, and shocked gas. On the other hand, Saken, Fesen \& Shull (1992) fitted only \textit{IRAS} infrared 
data with a two temperature model with (both) emissivity indices of 1.5, and estimated temperatures of the cold and hot dust components 
of 34.3 and 185~K, respectively. They used only the observed fluxes in the fitting procedure and not the color-corrected fluxes, since the 
color-correction factors are themselves model-dependent. It should be noted that there is a large disagreement regarding the true infrared 
flux densities in the literature, especially at 12 and 25 $\mu\mathrm{m}$ (see Table 6 in Saken et~al.\@ 1992).

In addition, spinning dust emission can shape the continuum spectra in a form of a characteristic bump between 10 -- 100~GHz 
(Erickson 1957; Draine \& Lazarian 1998a,b). It is an electric dipole emission from very small grains that spin rapidly due to the action of 
systematic torques in the interstellar medium. This emission mechanism is currently one of the most probable proposed mechanisms to explain 
the so called anomalous microwave emission (AME), i.e., dust-correlated emission from the Milky Way, observed between around 10 -- 100~GHz, 
that can not be accounted for by extrapolating the thermal dust emission to low frequencies (Planck Collaboration Int.\@ XV 2014, and references therein). 
The analysis of AME is of the great importance as studies of cosmic microwave background consider AME as an additional source of foreground contamination. 
Scaife et~al.\@ (2007) for the first time asserted the possibility of significant spinning dust emission from the vicinity of the SNR 3C 396. 
On the contrary, using new \textit{Parkes} 64-m telescope observations, Cruciani et~al.\@ (2016) found that presently known integrated continuum 
data do not favor the presence of either this emission component nor thermal bremsstrahlung radiation. Finally, based on the new {\it Planck's} data, 
Oni\'c (2015) proposed that the spinning dust mechanism can account for a significant excess emission at 30~GHz from the vicinity of SNR W44. 
Furthermore, G\'{e}nova-Santos et~al.\@ (2016) found a very compelling evidence for the spinning dust emission associated with W44.

The spinning dust emission is a very complex process. It actually depends on the size, shape, and charge of the emitting dust grains. This emission 
mechanism also depends on the environmental conditions such as gas temperature, molecular fraction, ionization state, and the intensity of the 
radiation field (Ali-Ha\"{\i}moud et~al.\@ 2009; Hoang, Draine \& Lazarian 2010; Ysard, Miville-Desch\^{e}nes \& Verstraete 2010; Ysard \& Verstraete 2010; 
Hoang, Lazarian \& Draine 2011; Silsbee, Ali-Ha\"{\i}moud \& Hirata 2011; Ysard, Juvela \& Verstraete 2011; 
Ali-Ha\"{\i}moud 2013; Hoang \& Lazarian 2016). The simple expression (that much simplifies the underlying physics) can be 
written as: \begin{equation} S_{\nu}^{\mathrm{Spd}}\propto\left({\nu}/{\nu_{0}}\right)^{2}{\exp\left[{1-\left({\nu}/{\nu_{0}}\right)^{2}}\right]},
\end{equation} where $\nu_{0}$ is the peak frequency usually between 10 and 70~GHz (Draine \& Hensley 2012; Hensley, Murphy \& Staguhn 2015). 
More robust analysis can be performed using the more advanced numerical schemes, e.g., \verb"SpDust" code, ver.\@ 2.01 
(Ali-Ha\"{\i}moud et~al.\@ 2009; Silsbee et~al.\@ 2011) or more realistic approximations such as of Stevenson (2014): \begin{equation} 
S_{\nu}^{\mathrm{Spd}}\propto\left({\nu}/{\nu_{0}}\right)^{c_{1}}\exp\left[-0.5\ \!c_{2} \ln^{2}({\nu}/{\nu_{0}})\right]\mathrm{erfc}\ \![c_{3}\ln({\nu}/{\nu_{0}})+c_{4}], 
\end{equation} where $c_{i}$ are particular model parameters (see Stevenson 2014 for details).

Hensley, Draine \& Meisner (2016) emphasized that the emission of spinning ultra-small grains might be enhanced in SNRs if shattering 
in grain-grain collisions increases the ultra-small grain population, or suppressed if ultra-small grains are destroyed 
by sputtering. In fact, several processes can excite or damp the grain's rotation: gas/grain interactions, photon emission (infrared and radio), 
formation of $\mathrm{H_{2}}$ molecules on the grain surface or photoelectric emission (Planck Collaboration XXI 2011). Moreover, 
the interstellar radiation field (ISRF) strongly affects the composition of the interstellar medium (ISM), since it can not only charge but 
also destroy dust grains when the intensity is high enough (Planck Collaboration Int.\@ XV 2014).

Finally, it is known that the integrated radio-continuum spectra of some dynamically evolved Galactic as well as of several 
extragalactic SNRs could appear in a (high-frequency) concave-down form (see Uro\v sevi\'c 2014 for a review). A few examples 
of Galactic SNRs with such radio spectra can be found in the recent literature, e.g., cases of SNRs S 147, HB 21, W44, IC 443, 
Puppis A,... (Xiao et~al.\@ 2008; Hewitt et~al.\@ 2012; Pivato et~al.\@ 2013; Planck Collaboration Int.\@ XXXI 2016). Within 
the sample of SNRs which spectra experience such spectral breaks there are examples that range from bright to faint and young to 
mature objects. It also includes SNRs with and without compact stellar remnants. Such spectral shapes may be caused by a 
combination of cosmic-ray acceleration by the shocks and the pulsars, deceleration in denser environments, as well as the aging 
(Leahy \& Roger 1998; Reynolds 2009; Planck Collaboration Int.\@ XXXI 2016, and references therein). In the case of some Large 
Magellanic Cloud SNRs, observed spectral breaks are explained by the effect of coupling between synchrotron losses and 
observational constraints (resolution) when the distant (extragalactic) emission region stayed unresolved (Bozzetto et~al.\@ 2010; 
de Horta et~al.\@ 2012; Bozzetto et~al.\@ 2012; Bozzetto et~al.\@ 2013).

Although they are just rough mathematical approximations, several simple expressions were used to fit those continuum spectra 
that manifest such spectral breaks. One such an expression is the simple smooth broken power-law: \begin{equation} 
S_{\nu}^{\mathrm{syn}}\propto(\nu/\nu_{\mathrm{b}})^{-\alpha_{1}}\left(1+(\nu/\nu_{\mathrm{b}})^{\Delta\alpha}\right)^{-1},\  \Delta\alpha=\alpha_{2}-\alpha_{1}>0, 
\end{equation} where $\alpha_{1}$ and $\alpha_{2}$ are particular spectral indices (for lower and higher frequency regimes) 
and $\nu_{\mathrm{b}}$ is the frequency at the spectral break. The simple exponential cut-off is another 
(see, e.g., Pivato et~al.\@ 2013): \begin{equation} S_{\nu}^{\mathrm{syn}}\propto\nu^{-\alpha}\exp[-\nu/\nu_{\mathrm{c}}], 
\end{equation} where $\nu_{\mathrm{c}}$ is the characteristic cut-off frequency.

It should be emphasized, at this point, that the observed synchrotron X-ray emission from the rims of several young supernova 
remnants allows us to study the high-energy tail of the energy distribution of electrons accelerated at the shock front. 
Although, synchrotron emission from the shock has not been detected in the X-ray observations of the SNR IC 443, it should be 
noted that, in general, the analysis of such non-thermal X-ray emission of mixed-morphology SNRs can provide information 
on the physical mechanisms that limit the energy achieved by the electrons in the acceleration process. Different physical 
mechanisms can be invoked to limit that maximum energy achieved by the electrons in the acceleration process, such as radiative 
losses, limited acceleration time available, and change in the availability of magneto-hydrodynamic waves above some wavelength 
(i.e., loss-limited, time-limited, and escape-limited scenarios; Reynolds 1998; Zirakashvili \& Aharonian 2007; Reynolds 2008; 
Vink 2012; Miceli et~al.\@ 2013, 2014; Pohl, Wilhelm \& Telezhinsky 2015). If synchrotron losses of electrons occur uniformly 
over the whole lifetime of the SNR, $t\ [10^{4}\ \mathrm{yr}]$, for magnetic field intensity $B\ [10\ \mu\mathrm{G}]$, the 
high-frequency turnover would happen at the cut-off frequency $\nu_{\mathrm{c}}\ [\mathrm{MHz}]=3.4\times10^{9}\ B^{-3}\ t^{-2}$ 
(Xiao et~al.\@ 2008). Finally, most recently, Auchettl et~al.\@ (2016) reported detection of non-thermal X-ray emission component 
from the molecular cloud interacting mixed-morphology SNR G346.6-0.2. They concluded, surprisingly, that this is most likely 
synchrotron emission produced by the particles accelerated at the shock.

\section{Analysis and results}
\label{sec:res}

The radio/microwave continuum data (the flux densities at different frequencies) for SNR IC 443 were taken from Table 2 of 
Castelletti et~al.\@ (2011), Table 1 of Gao et~al.\@ (2011), Table 3 of Reich et~al.\@ (2003) and Table 3 of 
Planck Collaboration Int.\@ XXXI (2016). The flux densities at frequencies higher than 408 MHz that could not been 
corrected to the scale of Baars et~al.\@ (1977), i.e., the correction factor was not available, were excluded from 
the analysis (see Table 2 in Castelletti et~al.\@ 2011). In addition, \textit{IRAS} infrared data point at 100 $\mu\mathrm{m}$, 
taken from Table 3 of Saken et~al.\@ (1992), was included in the analysis as it gives us information on the 'Wien side' of the 
thermal dust emission spectrum (although its associated relative error is around 14\% and there is a modest disagreement between 
the flux density values stated in the various literature -- see Table 6 from Saken et~al.\@ 1992).

The large scatter, even for data points at the same continuum frequencies, is apparent (see Fig.\@ 2 of this paper 
or Fig.\@ 7 from Castelletti et~al.\@ 2011). Generally, one should be very cautious in comparing data from different 
types of radio observations, e.g., single dish vs.\@ interferometer. One should ensure that the measured flux densities 
can be reliably compared. In fact, to draw a firm quantitative conclusions from such a combination of data (combined data sample) 
is very hard and can be rather biased. Another related drawback is that the flux densities at different frequencies were 
obtained from observations at different angular resolutions and using different techniques. In general, there could be 
completely different background contributions at different frequencies. One proposed strategy is to degrade 
all the maps to the same angular resolution, and use the same aperture size for all the frequencies as in G\'{e}nova-Santos et~al.\@ (2016) 
for the case of SNR W44. Nevertheless, we stay in a framework of qualitative interpretation and stress the fact that the integrated radio/microwave 
emission from SNR IC 443, as presented in the known literature, can not be explained solely by a combination of synchrotron and thermal dust emission.

To show that the observed high-frequency radio and microwave continuum of SNR IC 443 manifests genuine bump around 30~GHz, the principal component analysis 
(PCA) can be applied (Babu \& Feigelson 1996). The principal components are actually the eigenvectors of the covariance matrix. The original data can be 
represented in the (new) basis formed by these vectors. The role of the first principal component is that it accounts for as much of the variability in 
the data as possible. In fact, it has the largest possible variance. In other words, the eigenvector with the largest eigenvalue is the direction along 
which the data set has the maximum variance. The second principal component has the highest variance possible under the particular constraint of the 
orthogonality to the first one. Moreover, the PCA allows calculation of the highest data variability direction. That fact can be used to detect slight 
departures from the pure power-law spectra, i.e., linear function in $\log-\log$ scale.

In Fig.\@ 1, the results of PCA applied to logarithmically transformed original data ($\log\nu, \log S_{\nu}$) at frequencies between 400~MHz 
and 143~GHz are presented. These are high enough continuum frequencies such that the effects of the low-frequency spectral bending can be neglected. 
Abscissa (PC1) and ordinate (PC2) correspond to the first and second principal component directions, respectively. Plotted values are the (original) 
data represented in the (new) basis of principal components (zero centered data multiplied by the rotation matrix whose columns contain the eigenvectors). 
The covariance of these values represents a diagonal matrix with squares of standard deviations of the principal components eigenvalues as its elements. 
It is clear that the radio-spectrum follows pure power-law relation (linear relation in $\log-\log$ scale) in the range between around 400~MHz and 
5~GHz, although with undoubted scatter (open circles). Such a scatter is, in fact, expected due to the nature of the combined data sample as previously 
discussed. On the other hand, addition of the data points between 8 and 143~GHz (filled circles in Fig.\@ 1) shows that the continuum spectrum of IC 443 
slightly deviates from the simple power-law in that frequency range. Particularly, data point at 8~GHz, listed in Table 2 of Castelletti et~al.\@ (2011) 
and derived in the paper of Howard \& Dickel (1963) has a large relative error of 20\% (the first filled circle symbol to the left). In fact, 
that is not in favor of the genuine (physically originated) bump hypothesis. However, 5 {\it Planck} data points, between 30 and 143~GHz, have much 
smaller uncertainties (relative errors roughly between 5 and 8\%). Not only that the slight positive curvature (excess emission) around 30~GHz is 
apparent but also the slight concave-down feature (a dip in emissivity) at around 143~GHz is also clearly visible (filled circles in Fig.\@ 1). 
The latter can be easily explained by the change in the shape of the synchrotron spectrum as discussed in the end of the Section 4. As the continuum 
spectrum above 143~GHz is influenced by the thermal dust emission (deviation from the synchrotron power-law spectrum is a priori excepted), 
it was not of interest for the presented PCA analysis.

\begin{figure}[!t]
\figurenum{1}
\plotone{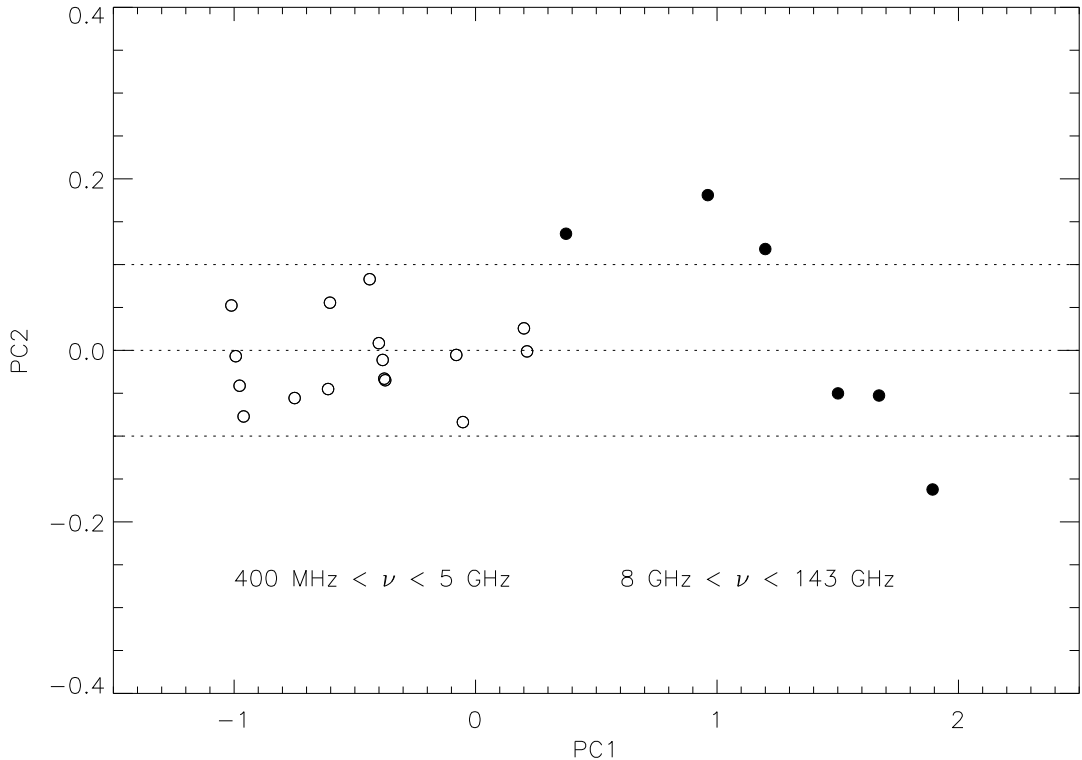}
\caption{The results of PCA on logarithmically transformed data ($\log\nu, \log S_{\nu}$) for SNR IC 443. Abscissa (PC1) and ordinate (PC2) 
correspond to the first and second principal component directions, respectively. Plotted values represent the (original) data in the basis 
of principal components (zero centered data multiplied by the rotation matrix whose columns contain the eigenvectors). The results of PCA 
on the data between 400~MHz and 5~GHz are marked with open circles. Slight departure from the linear relationship is obvious only when the data 
points between 8 and 143~GHz are included (filled circles).}
\label{fig:1}
\end{figure}

The SNR IC 443 surroundings is composed of a very large number of interacting environments, such as molecular clouds, warm ionized medium, warm neutral 
medium, etc. Due to the low quality of the overall radio/microwave spectrum of this remnant (large scatter in the combined data sample used in the analysis; 
significant number of data points with relative errors as high as 20\%) we confine ourselves to the qualitative analysis. In fact, distinguishing between 
different models is not at all trivial from currently available data. In that sense, to describe radio to microwave SED of the SNR IC 443, the following, 
simplified models were used: \begin{equation}S_{\nu}^{i}=S_{\nu}^{\mathrm{M}i}+S_{\nu}^{\mathrm{Spd}}+S_{\nu}^{\mathrm{Td}}(\beta_{\rm d},T_{\rm d}),
\end{equation} where $i$ stands for the models M1 (Model 1) and M2 (Model 2), defined by the Eq.\@ 2 and 3, respectively. We used Eq.\@ 8 to 
describe the high-frequency spectral bending. The thermal dust is fitted by the, oversimplified model that assumes one modified black-body 
(Planck-like) function (Eq.\@ 4) in accordance with Planck Collaboration Int.\@ XXXI (2016). We also use simple representation for the spinning 
dust emission, given by Eq.\@ 5, due to the impossibility of the firm breaking of the degeneracies between slightly different models.

The procedure of weighted least-squares fit is conducted by the \verb"MPFIT"\footnote{http://purl.com/net/mpfit} (Markwardt 2009) package written in 
\verb"IDL" throughout this paper, with starting values estimated from the data. We note that \verb"MPFIT" provides estimates of the 1$\sigma$ uncertainties 
for each parameter (the square root of the diagonal elements of the parameter covariance matrix).

\begin{figure}[!t]
\figurenum{2}
\plotone{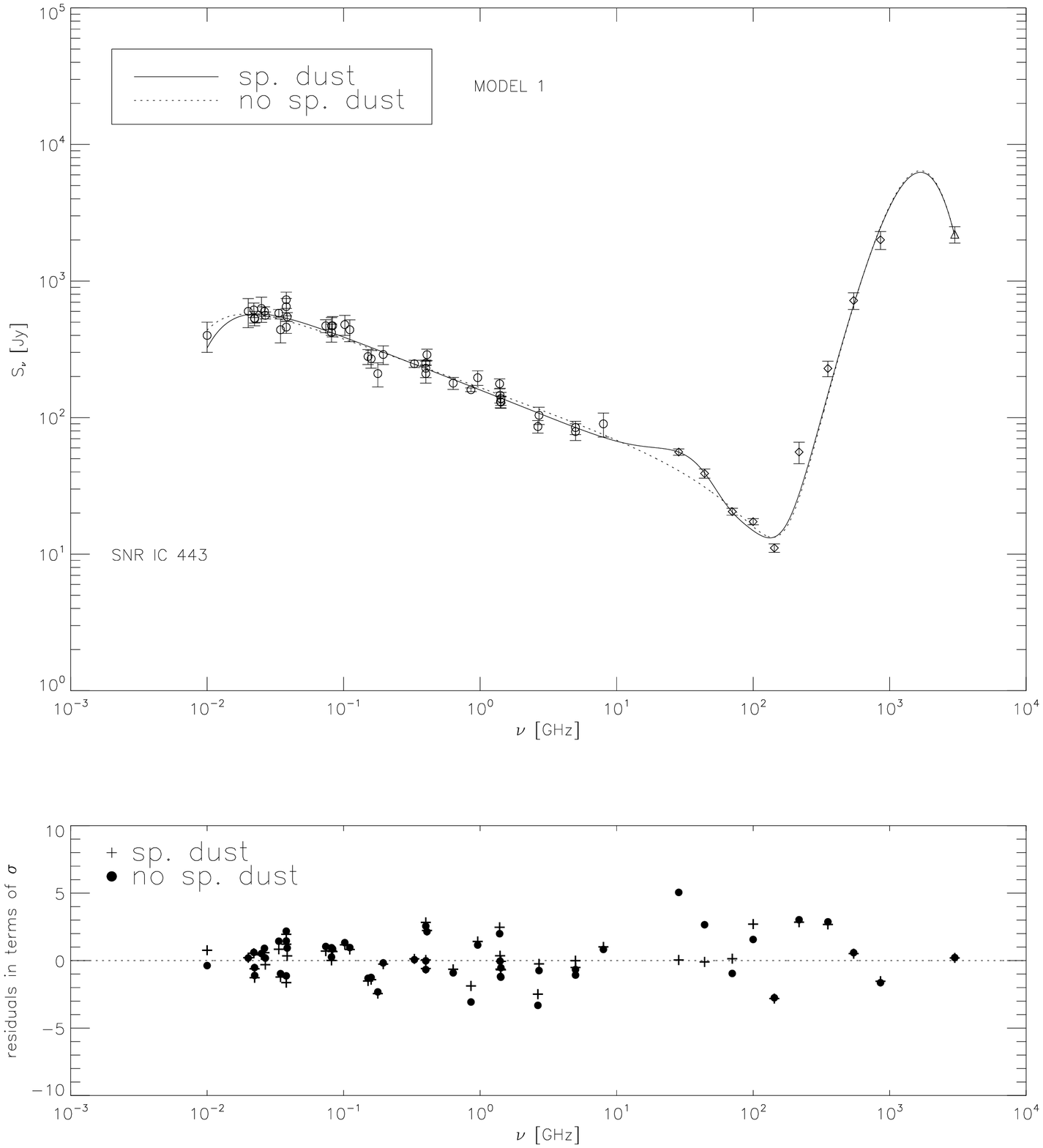}
\caption{The weighted least-squares fit to the data for the M1 model. Solid line represents the fit when spinning dust emission is included while 
dashed line correspond to the fit without spinning dust emission, made for a comparison. Diamond symbols indicate {\it Planck} data and \textit{IRAS} 
point at 100 $\mu\mathrm{m}$ is shown as a triangle. In the lower graph, residuals in terms of the particular data uncertainties ($\sigma$) are presented: 
plus symbols and filled circles hold for the model with and without spinning dust emission, respectively.}
\label{fig:2}
\end{figure}

\begin{figure}[!t]
\figurenum{3}
\plotone{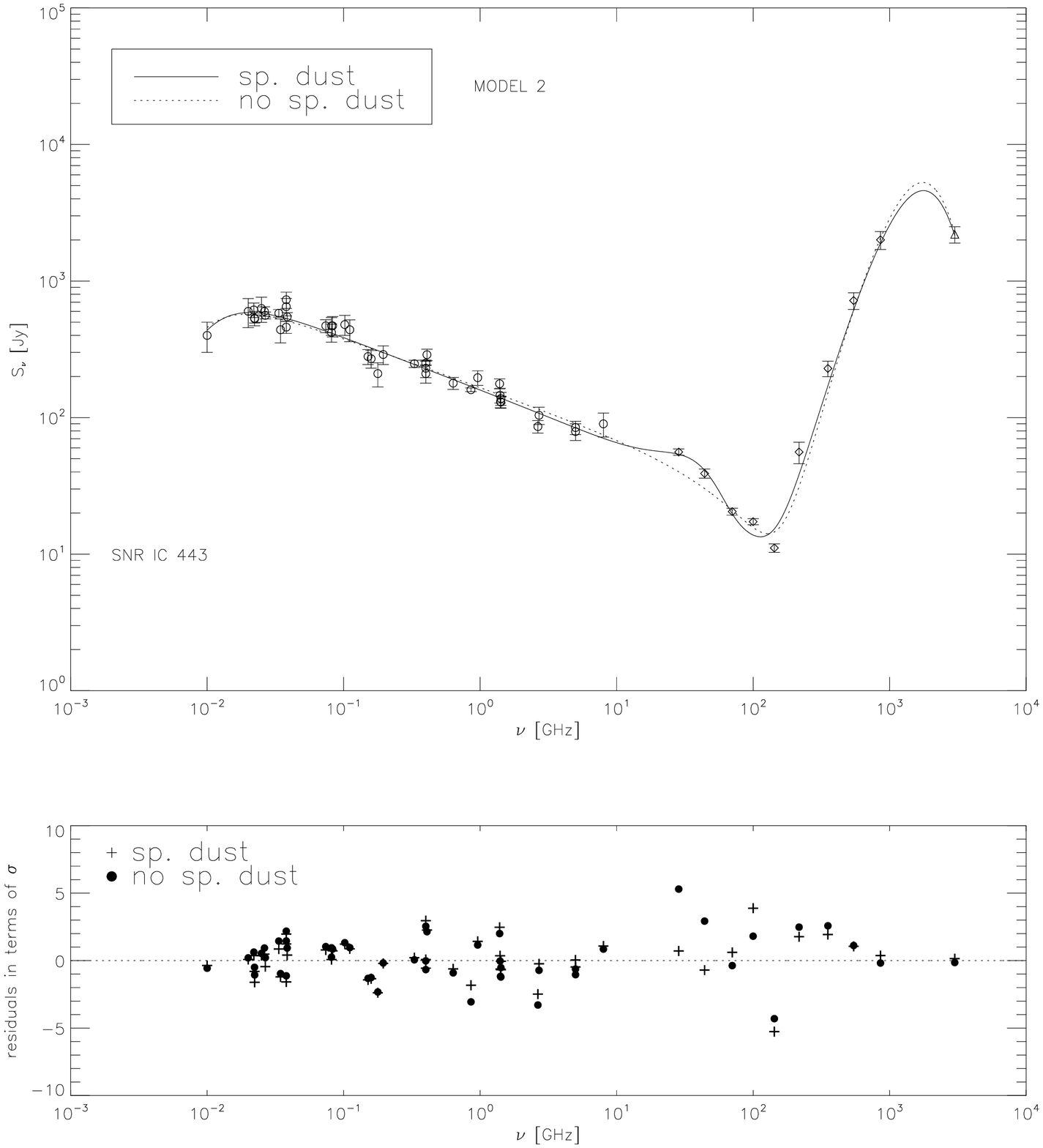}
\caption{The weighted least-squares fit to the data for the M2 model. Solid line represents the fit when spinning dust emission is included while 
dashed line correspond to the fit without spinning dust emission, made for a comparison. Diamond symbols indicate {\it Planck} data and \textit{IRAS} 
point at 100 $\mu\mathrm{m}$ is shown as a triangle. In the lower graph, residuals in terms of the particular data uncertainties ($\sigma$) are presented: 
plus symbols and filled circles hold for the model with and without spinning dust emission, respectively.}
\label{fig:2}
\end{figure}

In Figs.\@ 2/3, the weighted least-squares fit to the data, when M1/M2 models are applied is shown. Solid and dashed lines represent the fits with and without 
additional, spinning dust emission component, respectively. Diamond symbols indicate {\it Planck} data and \textit{IRAS} point at 100 $\mu\mathrm{m}$ 
is shown as a triangle. The data points with larger uncertainties have smaller weights which is clearly seen in the results (see Figs.\@ 2/3). In the 
lower graphs of Figs.\@ 2/3, residuals (the difference between the observed and model prediction value) in terms of the particular data uncertainties 
($\sigma$) are presented: plus symbols and filled circles hold for the model with and without spinning dust emission mechanism included, respectively. 
It is evident that the particular excess in flux density at around 30~GHz is very well explained by the addition of the spinning dust emission component. 
In fact, when spinning dust component is included, the data point at 28.5~GHz deviates only around $0.04\sigma$ (M1 model) and $0.7\sigma$ (M2 model) 
from the predicted values, which is very well inside the general scatter that is present in the data (see Figs.\@ 2/3). On the other hand, in the case 
of fit without any additional emission mechanism, the same data point deviates slightly more than $5\sigma$ from the predicted value (for both, M1 and M2 
model). However, data points at 100 and 143~GHz still significantly departure from the predicted values even in the case of fits with spinning 
dust emission included (their residuals are beyond the uppermost ones, i.e., those at 178~MHz and 408~MHz, with associated relative errors of around 
20\% and 10\%, respectively -- Bennett 1962; Colla 1971). The simple explanation can be found in the fact that, both, synchrotron cut-off as well as the 
thermal dust emission model should, in fact, be represented by more realistic (complicated) functions as discussed earlier. In addition, {\it Planck} 
data point at 217~GHz (with rather large relative uncertainty of around 18\%; Planck Collaboration Int.\@ XXXI 2016) also slightly deviates from the 
model predictions which is yet another indication that this part of the continuum should be represented by more advanced models (two temperature modified 
black-body thermal dust emission, more physically justifiable functions representing synchrotron high-frequency cut-off). As the uncertainty in 
the background subtraction generally makes an accurate flux density difficult to measure, the lack of good quality data, especially in the infrared range, 
permits such an analysis. Nevertheless, being just a contamination from a complex nearby regions or possibly, a genuine radiation from the SNR, 
spinning dust emission can very well account for the apparent excess emission in the currently presented integrated continuum of IC 443 at frequencies 
around 20 -- 70~GHz.

The best fitting parameters for analyzed models are presented in Table 1, particularly, radio synchrotron spectral index $\alpha$, 
(synchrotron) cut-off frequency $\nu_{\mathrm{c}}$, spinning dust fractions $f(30)$ (ratio between the spinning dust flux density estimate at 30~GHz, 
$S^\mathrm{Spd}(30)$, and total flux density at the same frequency: $f(30)=S^\mathrm{Spd}(30)/S(30)$), and peak frequency $\nu_{0}$, temperature 
of the thermal dust $T_{\rm d}$, emissivity index $\beta_{\rm d}$, and the values of $\chi^{2}$. In Table 1, parameter $k =  N - p$ 
(where $N$ is the number of data points and $p$ is the number of model parameters) is given for convenience. One should also bare in mind that 
in non-linear models, $k$ does not always represent the exact number of degrees of freedom which is generally unknown, i.e., it is not possible 
to compute the exact value of reduced $\chi^{2}$ (Andrae et~al.\@ 2010). Nevertheless, it is clear that the fits with M1 model have a bit smaller 
$\chi^{2}$ for the same $k$'s than those with M2 model.

It is worth noting that our model predictions for thermal dust parameters (taking the error estimates into account) are in general accordance 
with those for Galactic clouds (see, e.g., Table 3 from Planck Collaboration Int.\@ XV 2014). Still, M1 model predicts rather unusually high 
emissivity index $\beta_{\rm d}>2$, for both cases, with and without spinning dust emission (but see the discussion on radio-continuum of 
YSOs driving known outflows; AMI Consortium 2012), although other M1 model parameters do not deviate significantly from the M2 model results. 
Of course, it is highly probable that this discrepancy comes from the, already discussed, oversimplification of the continuum emission 
above around 70~GHz.

\begin{table}[h!]
\renewcommand{\thetable}{\arabic{table}}
\centering
\caption{The best fitting parameters for analyzed models.} \label{tab:1}
\begin{tabular}{cccccccc}
\tablewidth{0pt}
\hline
\hline
Model & $\alpha$ & $\nu_{\mathrm{c}}\ [\mathrm{GHz}]$ & $f(30)$ & $\nu_{0}\ [\mathrm{GHz}]$ & $T_{\rm d}\ [\mathrm{K}]$ & $\beta_{\rm d}$ & $\chi^{2}\ (k)$\\
\hline
M1 with sp.\@ dust & $0.39\pm0.01$ & $152\pm24$ & $0.36$ & $28\pm3$ & $15.2\pm0.9$ & $2.34\pm0.23$ & $95\ (42)$\\
M1 without sp.\@ dust & $0.35\pm0.01$ & $129\pm15$ & \nodata & \nodata & $14.9\pm0.9$ & $2.43\pm0.23$ & $138\ (44)$\\
M2 with sp.\@ dust & $0.38\pm0.01$ & $112\pm16$ & $0.37$ & $32\pm3$ & $18.5\pm1.4$ & $1.64\pm0.21$ & $115\ (42)$\\
M2 without sp.\@ dust & $0.36\pm0.01$ & $120\pm13$ & \nodata & \nodata & $17.0\pm1.4$ & $1.97\pm0.30$ & $146\ (44)$\\
\hline
\end{tabular}
\end{table}

Our fits also do not favor significant contribution of thermal bremsstrahlung emission component. In addition, if we exclude the infrared data point 
at 100 $\mu\mathrm{m}$, the fits degrade in quality although the results do not change significantly from the presented ones. Finally, the rather 
large values of $\chi^{2}$ (see Table 1) are not at all surprising keeping in mind (1) the large scatter in data points, (2) simplified models used.

The radio-synchrotron spectral index agrees well with the results of Planck Collaboration Int.\@ XXXI (2016) although they did not assume the 
low-frequency bending (thoroughly discussed in Castelletti et~al.\@ 2011) as they did not fit the entire radio continuum (from the lowest 
frequency measurement at 10~MHz with a rather high relative uncertainty of 25\%; Bridle \& Purton 1968). Actually, they used the simple 
broken power-law as a model of synchrotron radiation in addition to the one modified black-body (MBB) component to represent the thermal dust 
emission. For a comparison, we modeled our data sample with smooth broken power-law (Eq.\@ 7) and one MBB component (Eq.\@ 4). Although the 
fit is not of significantly less quality ($\chi^{2}=132,\ k=45$) than the results presented in Table 1, the high-frequency spectral index 
$\alpha_{2}$ has a rather high value of around 3 ($\alpha_{1}=0.34\pm0.01,\ \Delta\alpha=2.70\pm0.48$,\ $\nu_{\mathrm{b}}=84\pm4$). 
This is much higher than the values considered in theory that supports the proposition that the breaks in spectral index are consistent 
with synchrotron losses of electrons injected by a central source (see, e.g., Reynolds 2009). This model also does not take into account 
the low-frequency thermal absorption effects.

We should also note that fits without high-frequency synchrotron spectral steepening are not favored. However, using our data sample we 
can not thoroughly disentangle between the synchrotron models that incorporate smooth broken power-law and exponential cut-off.

Finally, we would also like to point out that there is a significant difference between the data sample used by Planck Collaboration Int.\@ XXXI (2016) 
and the one presented in this paper. Former used only the data above 1~GHz (in fact, only 6 data points taken from Table 2 of Castelletti et~al.\@ (2011) 
in addition to the {\it Planck} measurements). They also used the data for which a correction factor to the scale of Baars et~al.\@ (1977) was unavailable, 
e.g., they included the 10.7~GHz point from Kundu \& Velusamy (1972). These differences are important as may change the conclusions significantly.

\section{Discussion}
\label{sec:disc}

Additional support for the spinning dust emission hypothesis can be found in certain correlations with thermal dust emission, 
especially at \textit{IRAS} wavelengths. Particularly, Planck Collaboration Int.\@ XV (2014) found that, usually, for 
a spinning dust emission sources, a $12\ \mu\mathrm{m}$/$25\ \mu\mathrm{m}$ intensity ratio is $\approx(0.6-1.0)$. For various 
observations of IC 443 at these wavelengths (see Table 6 from Saken et~al.\@ 1992), it leads to around 0.6 -- 0.9 for the SNR IC 443, 
which is in the agreement with above predictions. However, ionic line contamination of the particular \textit{IRAS} bands, 
as emphasized in Oliva et~al.\@ (1999), makes this correlation uncertain. Another known indicator is the ratio between 
the spinning dust flux density estimate at 30~GHz and the $100\ \mu\mathrm{m}$ (3000~GHz) flux density which is usually around 
$(1-15)\times10^{-4}$ (Planck Collaboration Int.\@ XV 2014; Hensley et~al.\@ 2016). Bearing in mind the roughness of 
our analysis, we obtained larger values in range of $(90-100)\times10^{-4}$, for both models used in this paper and various 
infrared observations from the literature. In addition, spinning dust fractions are, for both models, slightly less than 
$f(30)\approx0.5$ (see Table 1), which was found in Planck Collaboration Int.\@ XV (2014) for the known AME regions.

One should also consider the possible contribution of ultra-compact \mbox{H\,{\sc ii}} regions (\mbox{UCH\,{\sc ii}}) to the detection 
of diffuse AME. At low frequencies, typically under 15~GHz, these objects may be optically thick, moving into the optically thin 
regime at frequencies higher than 15~GHz (Irfan et~al.\@ 2015). If an \mbox{UCH\,{\sc ii}} region, optically thick at 5~GHz, is 
positioned within the observed source region this would result in what would appear to be excess emission at higher frequencies where the 
\mbox{UCH\,{\sc ii}} region becomes optically thin. To ensure that the excess emission seen around 30~GHz is due to the spinning dust, 
the source region must be checked for nearby optically thick \mbox{UCH\,{\sc ii}} regions. We used \textit{IRAS} Point Source Catalog 
(Beichman et al.\@ 1988) to identify such objects in $1\degr$ radius region around the established central coordinates of IC 443. 
In fact, \mbox{UCH\,{\sc ii}} regions generally possess \textit{IRAS} color ratios of $\log(S_{60\ \!\! {\rm \mu m}}/S_{12\ \!\! {\rm \mu m}})\geqslant1.30$ 
and $\log(S_{25\ \!\! {\rm \mu m}}/S_{12\ \!\! {\rm \mu m}})\geqslant0.57$, which can be used as an identification method for these objects 
(Wood \& Churchwell 1989; Dickinson 2013; Planck Collaboration Int.\@ XV 2014). Extragalactic sources as well as those with only upper 
limits for the \textit{IRAS} fluxes listed, were excluded. No sources, matching all the criteria were found.

We want to emphasize the importance of discrimination between the free-free emission from YSOs and spinning dust emission (Scaife 2012). 
YSOs can generate emission at cm wavelengths due to a variety of mechanisms such as stellar winds and/or shock-induced ionization, which 
can mimic the spinning dust emission spectral signal (AMI Consortium 2010; Tibbs et~al.\@ 2015). In fact, understanding if the observed sources 
are both simultaneously forming stars and harboring spinning dust emission can help to understand the potential role of spinning dust emission in 
the star formation process. As YSOs are identified in the IC 443 region by the infrared observations, further, high-resolution cm 
observations of that area would be of the great importance for this study. Still, we note that any additional free--free component to our models 
(excluding spinning dust component) is not favored. Of course, due to the low quality of presently known IC 443 radio/microwave continuum 
we can not make any definite conclusions.

In their study of AME in Galactic clouds, Planck Collaboration Int.\@ XV (2014) did not list IC 443 area as a candidate AME emission region 
(such as, e.g., W48 region containing SNR W44). On the other hand, Planck Collaboration X (2016) decomposed the full-mission all-sky 
\textit{Planck} observations (Planck Collaboration I 2016) into several foreground components (i.e., synchrotron, free-free, AME,...) making use 
of both the 9 year \textit{WMAP} data (Bennett et~al.\@ 2013) and the Haslam 408~MHz survey (Haslam et~al.\@ 1982). In that sense, among the 
others, a full-sky map of the AME emission component was produced. Spanning a range of 408~MHz to 857~GHz in frequency Planck Collaboration X (2016) 
actually performed foreground component separation within the Bayesian \verb"Commander" analysis framework (Eriksen et~al.\@ 2004, 2006, 2008). 
They modeled AME component by the sum of two spinning dust spectra with fixed spectral shape as determined by the \verb"SpDust" code, but differing 
amplitudes and peak frequencies. In this phenomenological model, one of the spectra was required to have a spatially fixed peak frequency, fit to 
be 33.35~GHz, while the other peak frequency was allowed to freely vary from pixel to pixel. In that sense, it is worth checking the resulting AME 
foreground map from the \textit{Planck} Legacy Archive\footnote{http://pla.esac.esa.int/pla} for any significant emission in direction 
of the SNR IC 443. AME map has an angular resolution of $1\degr$ full-width half-maximum (FWHM) with 
\verb"HEALPix"\footnote{http://healpix.sourceforge.net/} resolution of $\mathrm{N}_{\rm side}=256$ (G\'{o}rski et~al.\@ 2005). Of course, 
one should bare in mind that the (mean and usually quoted) angular radius of SNR is around $22\farcm5$ (Green 2014).

Adopting the central radio coordinates of IC 443 from Green (2014), we measured counts in circular regions with $30\arcmin$ and $1\degr$ radii 
for both AME components. Using Rayleigh-Jeans formula we can roughly estimate the flux density from these regions (baring in mind pixel dimensions 
of $21\arcmin\times21\arcmin$). For AME1 component (with reference frequency of 22.8~GHz) it leads to around 6~Jy and 19~Jy, respectively. 
The results for AME2 component (with reference frequency of 41.0~GHz and spinning dust peak frequency of 33.35~GHz) are around 4~Jy and 11~Jy, 
for the analyzed circular regions, respectively. However, our model predictions for $S^\mathrm{Spd}$, from the analysis of the 
integrated radio to microwave continuum spectrum of IC 443, are a bit larger (even when we sum both AME components at the same frequency): 
$18\pm6$~Jy and $16\pm4$~Jy at 22.8~GHz, and $14\pm8$~Jy and $17\pm7$~Jy at 41.0~GHz, for M1 and M2 model, respectively. In addition, 
due to the low resolution, these estimated AME \verb"Commander" flux density values do not represent only the emission from the SNR, 
but also sample the emission from complex nearby regions (i.e., Galactic clouds that can be places of significant spinning dust emission; 
Planck Collaboration Int.\@ XV 2014).

As noted by Planck Collaboration X (2016), although the \verb"Commander" AME map provides a good tracer of spinning dust in our Galaxy, 
however, there are significant degeneracies between the free-free and AME components. That fact should be taken into consideration 
when using these foreground maps to estimate emissivities. Among the others, there is a possibility of significant leakage between 
synchrotron, AME (spinning dust), and free-free components. Such a leakage can generally be the source of a decrement in the \verb"Commander" 
synchrotron maps (see, e.g., Planck Collaboration XXV 2016). We would like to emphasize that very important drawback of the \verb"Commander" 
fits is related to the particular synchrotron model. It is very simple and perhaps not even suitable for an SNR (see Planck Collaboration X 2016, 
and references therein for the definition of that model). Furthermore, the \verb"Commander" synchrotron flux density at 408~MHz is around 92~Jy 
for the measured counts in circular region with $30\arcmin$ radius around the central position of IC 443. The estimates of flux density 
at the same frequency from our model fits are roughly around 225~Jy. At that particular continuum frequency the synchrotron component 
should dominate other proposed emission mechanisms for this SNR.

Baring in mind the possibility of component separation as well as particular model related issues, it is worth noting that currently known 
AME foreground map, at least, does not rule out our hypothesis that spinning dust emission in the SNR IC 443 region can be significant 
enough to influence presently known integrated radio/microwave continuum of this remnant.

In the end, it is worth mentioning that the {\it LOFAR} ({\it Low-Frequency Array}) allows detailed sensitive high-resolution studies of the 
low-frequency (10 -- 240~MHz) radio sky (van Haarlem et~al.\@ 2013). With a possible baseline length of around one thousand kilometers, 
the angular resolution of {\it LOFAR} extends to sub-arcsecond scales. In that sense, it can be used to shed light on the physical origin 
of the low-frequency turnover in the integrated continuum radio-spectrum of SNR IC 443. Complementary to the lowest radio-frequencies, 
more data in the frequency range between 10 and 100~GHz, at much higher angular resolution than the one of {\it Planck}, are needed 
to make firm conclusions about the contribution of particular radiation mechanisms responsible for an observed shape of the integrated 
radio/microwave continuum of this remnant. {\it ALMA} ({\it The Atacama Large Millimeter/Submillimeter Array}), for example, will, 
upon completion, cover continuum frequencies in 31 -- 950~GHz range (ALMA Partnership 2016). Depending on particular configuration 
and frequency band, the angular resolution will range from several arcseconds to the order of milliarcseconds. In addition, 
observations by {\it S-PASS} ({\it S-Band Polarization All Sky Survey}) at 2.3~GHz and angular resolution of around $9\arcmin$, as well as 
observations by {\it C-BASS} ({\it The C-Band All Sky Survey}) at 5~GHz and {\it QUIJOTE} ({\it Q-U-I JOint Tenerife CMB Experiment}) at 10--40~GHz 
with rather poor angular resolution, of around $1\degr$, will also help to considerably improve our knowledge on the continuum spectrum of several 
Galactic SNRs (Carretti 2011; King et~al.\@ 2014; Rubi\~{n}o-Mart\'{\i}n et~al.\@ 2012).

\section{Conclusions}

In this work, different emission models that can be responsible for the particular shape of the integrated radio/microwave 
continuum spectrum of Galactic supernova remnant IC 443 are tested and discussed. 

\begin{enumerate}
\item Recent observations by \textit{Planck} space telescope make it possible to analyze the high-frequency part of radio/microwave emission from the SNRs.

\item The possibility that the slight bump in the integrated continuum of this remnant around 20 -- 70~GHz is genuine and that can be explained by the 
contribution of additional emission mechanism such as of spinning dust is emphasized. In fact, even considering all the drawbacks of the presented analysis, 
the quality of the fit is significantly improved when spinning dust emission is included in the spectral model. In addition, the \verb"Commander" AME 
foreground map does not rule out the possibility of significant spinning dust emission from the IC 443 region. Finally, models that include the 
high-frequency synchrotron bending of the IC 443 radio-spectrum are favored.

\item New {\it LOFAR} data will presumably shed light on the physical origin of the low-frequency turnover in the integrated continuum radio-spectrum of SNR IC 443. 
Complementary to the lowest radio-frequencies, more data in the frequency range between 10 and 100~GHz, at much higher angular resolution than the one of {\it Planck}, 
are needed to make firm conclusions about the contribution of particular radiation mechanisms responsible for an observed shape of the integrated radio/microwave 
continuum of this remnant. 
\end{enumerate}

\acknowledgements{We would like to thank the anonymous referee for valuable suggestions that substantially improved this paper. In addition we would like 
to thank B.\@ Arbutina and B.\@ Vukoti\' c for helpful comments. This work is a part of Project No.\@ 176005 "Emission nebulae: structure and evolution" 
supported by the Ministry of Education, Science, and Technological Development of the Republic of Serbia. This work is also partially supported by 
a grant from the Natural Sciences and Engineering Research Council of Canada.}



\end{document}